\documentclass[journal,twoside,print]{ieeecolor2}
\usepackage{generic}
\usepackage{cite}
\makeatletter
\let\NAT@parse\undefined
\makeatother
\usepackage{amsmath,amssymb,amsfonts}
\usepackage{algorithmic}
\usepackage{graphicx}
\usepackage{algorithm,algorithmic}
\usepackage[hidelinks]{hyperref}
\usepackage{textcomp}
\usepackage[acronym]{glossaries}
\usepackage{booktabs} 
\usepackage{multirow}
\usepackage{bm}
\usepackage{subcaption}
\newacronym{sEMG}{sEMG}{surface electromyography}
\newacronym{PINN}{PINN}{physics-informed neural network}
\newacronym{PENN}{PENN}{Physics-Embedded Neural Network}
\newacronym{MSK}{MSK}{musculoskeletal}
\newacronym{CNN}{CNN}{convolutional neural network}
\newacronym{NRMSE}{NRMSE}{normalized root mean square error}
\newacronym{R2}{$R^2$}{coefficient of determination}
\newacronym{LSTM}{LSTM}{Long Short-Term Memory}
\newacronym{Bi-LSTM}{Bi-LSTM}{Bidirectional LSTM}
\newacronym{DoF}{DoF}{degrees of freedom}
\newacronym{NMF}{NMF}{nonnegative matrix factorization}
\newacronym{APL}{APL}{abductor pollicis longus}
\newacronym{FE}{FE}{Flexion/Extension}
\newacronym{RUD}{RUD}{Radial/Ulnar Deviation}
\newacronym{PCA}{PCA}{principal component analysis}
\newacronym{ICA}{ICA}{independent component analysis}
\newacronym{MVC}{MVC}{maximum voluntary contraction}
\newacronym{MSE}{MSE}{mean squared error}
\newacronym{CW}{CW}{clockwise circumduction}
\newacronym{CCW}{CCW}{counterclockwise circumduction}
\newacronym{CC}{CC}{Pearson correlation coefficient}
\newacronym{SC}{SC}{Spearman rank correlation coefficient}
\newacronym{FCR}{FCR}{flexor carpi radialis}
\newacronym{FCU}{FCU}{flexor carpi ulnaris}
\newacronym{ECRL}{ECRL}{extensor carpi radialis longus}
\newacronym{ECRB}{ECRB}{extensor carpi radialis brevis}
\newacronym{ECU}{ECU}{extensor carpi ulnaris}
\newacronym{TCN}{TCN}{temporal convolutional network}
\newacronym{PET}{PET}{parallel efficient transformer}
\newacronym{RND}{RND}{random motion}
\newacronym{VAF}{VAF}{variance accounted for}

\def\BibTeX{{\rm B\kern-.05em{\sc i\kern-.025em b}\kern-.08em
    T\kern-.1667em\lower.7ex\hbox{E}\kern-.125emX}}
\makeatletter
\def\arxiv@header{%
  \vbox{%
    \hbox to\textwidth{\hfill\scriptsize\textsf{\thepage}}%
    \vskip1pt%
    {\color{subsectioncolor}\hrule height1pt width\textwidth depth0pt}%
  }%
}
\def\ps@headings{%
  \def\@oddhead{\arxiv@header}%
  \def\@evenhead{\arxiv@header}%
  \def\@oddfoot{}%
  \def\@evenfoot{}%
}
\def\ps@titlepagestyle{%
  \def\@oddhead{\arxiv@header}%
  \def\@evenhead{\arxiv@header}%
  \def\@oddfoot{}%
  \def\@evenfoot{}%
}
\makeatother
\begin{document}
\bstctlcite{IEEEexample:BSTcontrol}
\title{Physiologically Constrained  Musculoskeletal Neural Network for Multi-DoF Joint Kinematics Estimation from Partially Observed sEMG}
\author{
Wending Heng$^{1}$,
Mingming Zhang$^{2}$,
Glen Cooper$^{1}$,
and Zhenhong Li$^{1,\dag}$
\vspace{1mm}\\
$^{1}$University of Manchester\quad
$^{2}$Southern University of Science and Technology\quad
\thanks{$^{\dag}$ Corresponding Author (email:zhenhong.li@manchester.ac.uk)}
}
\maketitle

\begin{abstract}
This paper investigates multi-\gls{DoF} joint kinematics estimation under partially observed \gls{sEMG}, where only a subset of task-relevant muscles can be measured due to anatomical inaccessibility or sensor constraints. A novel musculoskeletal neural network (\acrshort{MSK}-NN) is proposed to estimate multi-\gls{DoF} joint angles while simultaneously inferring activations for both measured and unmeasured muscles.
\acrshort{MSK}-NN consists of a \acrshort{CNN}-based muscle activation estimator and an embedded \acrshort{MSK} forward dynamics module, forming a fully differentiable architecture. 
Unlike existing hybrid neural frameworks that require additional biomechanical labels (e.g., muscle-tendon forces, joint torques), \acrshort{MSK}-NN is trained without direct supervision of internal biomechanical variables. A composite physics-physiology loss is designed by incorporating a joint kinematics loss, a data-driven muscle synergy loss, and an anatomy-guided trend loss.
The proposed method is evaluated on two-\gls{DoF} wrist kinematics estimation across three rhythmic motions with unconstrained speed and amplitude, and one random motion. Compared with \acrshort{CNN}, \acrshort{Bi-LSTM}, \acrshort{CNN}-LSTM, and \acrshort{PET} baselines, \acrshort{MSK}-NN achieves lower \gls{NRMSE} and higher \gls{R2}, especially for the random motion. More importantly, the optimized \acrshort{MSK} parameters remain within physiological limits, and the estimated activation of an input-excluded muscle exhibits strong temporal agreement with its recorded \gls{sEMG} envelope, demonstrating the capability of \gls{MSK}-NN to recover physiologically plausible activations. 
\end{abstract}

\glsresetall

\section{Introduction}
\label{sec:introduction}
\glsunset{sEMG}
\IEEEPARstart{S}{urface} electromyography (\gls{sEMG})-based continuous joint kinematics estimation has been widely used to decode human motion intention in rehabilitation \cite{weiContinuousMotionIntention2024d,zhaoAdaptiveCooperativeControl2023c,zhengAdaptiveComplianceControl2025, campaniniSurfaceEMGClinical2020}, biomechanics analysis\cite{Esrafilian2025EMGAssistedMSK,bermanEfficientFrameworkPersonalizing2024}, and human-robot interaction \cite{Cimolato2022,jiangBioroboticsResearchNoninvasive2023b, Zeng2021sEMGStiffness, zhangElectromyographySignalsbasedHumanrobot2022e}.

Model-free approaches employ data-driven regressors to learn a direct mapping from \gls{sEMG} to joint kinematics without explicitly modeling the underlying neuromuscular and biomechanical mechanisms (i.e., \gls{MSK} forward dynamics) \cite{maBiDirectionalLSTMNetwork2021b, baoCNNLSTMHybridModel2021b, linParallelEfficientTransformer2025}. Although these methods may achieve high kinematic estimation accuracy under training-like conditions, they often lack physiological interpretability and may produce biomechanically inconsistent predictions for movement patterns with speeds or amplitudes different from those seen during training \cite{Xiong2021EMGReview,weiContinuousMotionIntention2024d}.
In contrast, model-based approaches explicitly represent the \gls{MSK} forward-dynamics chain through activation dynamics, muscle--tendon dynamics, and joint dynamics. Beyond joint kinematics, they also provide intermediate variables (e.g., muscle activations, muscle forces, and joint torques) that are valuable for biomechanical analysis, clinical interpretation, and rehabilitation assessment.
However, their performance is often limited by the fidelity of the underlying \gls{MSK} model \cite{buchananNeuromusculoskeletalModelingEstimation2004, zhaoComputationallyEfficientPersonalized2023a}. Unmeasured muscles, such as deep or small muscles inaccessible to \gls{sEMG} recordings, are often omitted even though they may contribute substantially to joint torque and stability. This incomplete muscle-drive representation can lead to physiologically implausible force sharing and degraded kinematic estimation accuracy. Moreover, the uniform activation dynamics commonly used in model-based approaches do not fully capture heterogeneous neuromuscular characteristics across muscles. 
\begin{figure*}[!t]
  \centering
  \includegraphics[width=1\textwidth]{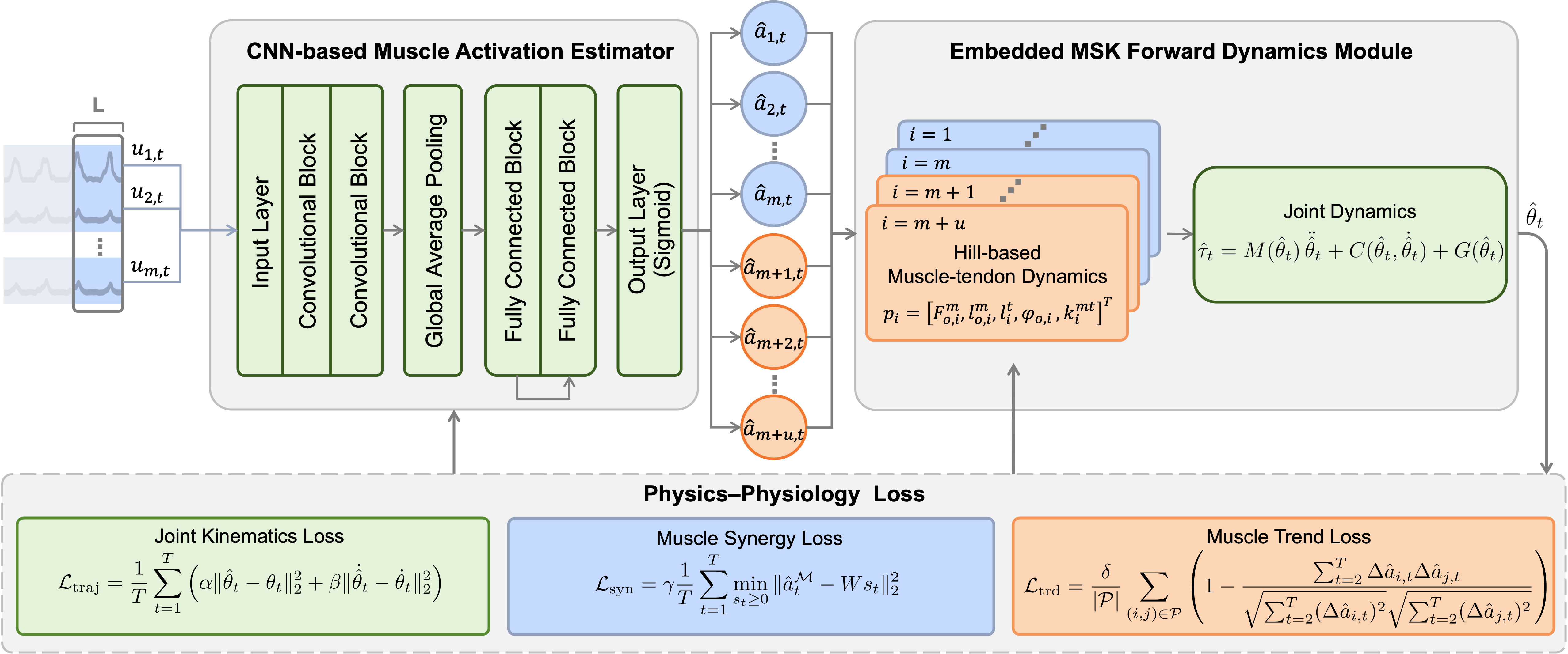} 
    \caption{Overview of the proposed \gls{MSK}-NN framework. It is composed of a Muscle Activation Estimator and an Embedded \gls{MSK} Forward Dynamics Module. The estimator takes the enveloped \gls{sEMG} from measured muscles $u_t = [u_{i,t}]_{i=1}^{n_m}$ as input and jointly estimates activations of both measured muscles (shown in blue) and unmeasured muscles (shown in orange), $\hat{a}_t = [\hat{a}_{i,t}]_{i=1}^{n_m+n_u}$. 
  A Physics-Physiology Embedded Loss jointly constrains the Muscle Activation Estimator and the muscles parameters $p_i$, enforcing consistency across kinematics, measured-muscle synergies, and unmeasured-muscle trends.}
  \label{fig:penn-framework}
\end{figure*}
To improve physiological and biomechanical consistency, hybrid neural frameworks have been proposed by combining data-driven learning with prior neuromusculoskeletal knowledge.
\Gls{PINN} methods incorporate joint dynamics into the loss function as soft constraints, guiding predicted muscle--tendon forces and joint kinematics toward prescribed physical relationships \cite{zhangPhysicsInformedDeepLearning2023c,shiPhysicsInformedLowShot2024,kumarPredictingMultiJointKinematics2023,zhangBoostingPersonalizedMusculoskeletal2023b}. However, soft constraints cannot always be guaranteed, especially when they compete with other objectives during training. To address this limitation, \gls{PENN} embeds \gls{MSK} forward dynamics into the network architecture to enforce physical consistency and uses a residual network to correct the remaining kinematic discrepancy, achieving high kinematic estimation accuracy \cite{hengPhysicsEmbeddedNeuralNetworks2025}. Nevertheless, existing hybrid neural frameworks either cannot provide physiologically plausible intermediate variables or require expensive biomechanical labels (e.g., muscle-tendon forces or inverse-dynamics joint torques) for supervision, which limits their scalability. More importantly, they assume a one-to-one mapping between \gls{sEMG} channels and modeled muscles, as in conventional model-based approaches. This assumption excludes unmeasured muscles and can lead to physiologically implausible force sharing between muscles.

In this paper, we investigate multi-\gls{DoF} joint kinematics estimation under partially observed \gls{sEMG}, where some task-relevant muscles are unmeasurable because of anatomical inaccessibility or sensor constraints (e.g., deep or small muscles). A novel \gls{MSK}-NN framework is proposed to estimate the joint kinematics while inferring physiologically plausible activations for both measured and unmeasured muscles. To improve interpretability, \gls{MSK}-NN adopts an architecture similar to model-based approaches, consisting of a muscle activation estimator and an \gls{MSK} forward-dynamics module composed of embedded Hill-based muscle-tendon dynamics and joint dynamics. To remove the requirement for one-to-one \gls{sEMG}-to-muscle mapping, the activation estimator uses a \gls{CNN}-based encoder to jointly estimate activations for measured and unmeasured muscles, allowing task-relevant unobserved muscles to contribute to kinematics estimation through physiologically structured torque generation. Inspired by prior \gls{sEMG} generation studies that use muscle synergy and co-contraction patterns of functional grouped muscles to reconstruct \gls{sEMG} in a low-dimensional space \cite{treschMatrixFactorizationAlgorithms2006,biancoCanMeasuredSynergy2018,aoEMGdrivenMusculoskeletalModel2022a,aoComparisonSynergyExtrapolation2024a, rabbiMuscleSynergyInformed2024}, we design a composite physics-physiology loss to regularize the underdetermined latent muscle activation estimation. This loss includes a joint kinematics loss on angle and angular velocity, a muscle-synergy loss on measured muscles, and an anatomy-guided trend  loss linking unmeasured muscles with functionally related measured muscles to encourage shared temporal coordination structure.  \gls{MSK}-NN is fully differentiable, and parameters of \gls{CNN}-based encoder and parameters of Hill-based muscle-tendon dynamics are optimized through end-to-end backpropagation.  The framework is evaluated on two-\gls{DoF} wrist kinematics estimation across three rhythmic motions at unconstrained speeds and amplitudes, plus a random motion, and demonstrates superior performance in \gls{NRMSE} and \gls{R2} compared with \gls{Bi-LSTM} \cite{maBiDirectionalLSTMNetwork2021b}, \gls{CNN}-LSTM \cite{baoCNNLSTMHybridModel2021b}, and \gls{PET} \cite{linParallelEfficientTransformer2025} baselines, particularly under the random motion (demonstrating strong generalizability). More importantly, the estimated activation of an input-excluded muscle shows strong temporal agreement with its recorded \gls{sEMG} envelope, demonstrating the capability of \gls{MSK}-NN to provide physiologically plausible muscle activations from partially observed \gls{sEMG}. Muscle ablation analysis further shows progressive performance degradation as modeled unmeasured muscles are removed, confirming their importance for  kinematics estimation.

\section{Methodology}
\subsection{Problem Formulation}
Consider the rotational motion of a single joint with $d$-\gls{DoF} driven by $n$ muscle–tendon units.
The index set of muscles with measurable \gls{sEMG} is denoted by $\mathcal{M}_m = \{1, \ldots, n_m\}$, where $|\mathcal{M}_m|=n_m$. Muscles whose \gls{sEMG} is unmeasurable due to anatomical inaccessibility or sensor constraints (e.g., deep or small muscles, and accessible muscles that are not measured because of limited sensor availability) are denoted by $\mathcal{M}_u = \{n_m+1, \ldots, n_m+n_u\}$, where $|\mathcal{M}_u|=n_u$. The complete muscle set is $\mathcal{M} = \mathcal{M}_m\cup\mathcal{M}_u$, with $|\mathcal{M}|=n$.

The objective is to estimate the joint angle ${\theta} \in \mathbb{R}^d$ based on \gls{sEMG} measured over muscles in $\mathcal{M}_m$, while estimating activations for muscles in $\mathcal{M}$. This task requires multi-\gls{DoF} kinematics estimation under partially observed \gls{sEMG} while providing physiologically plausible activations for both measurable and unmeasurable muscles.

\subsection{\gls{MSK}-NN Framework}
The proposed \gls{MSK}-NN framework consists of an activation estimator and a \gls{MSK} forward dynamics module, as shown in Fig.~\ref{fig:penn-framework}.
The estimator takes pre-processed \gls{sEMG} from measurable muscles as input and estimates the activations of both measurable and unmeasurable muscles. These activations are propagated through Hill-type muscle-tendon dynamics models \cite{thelenAdjustmentMuscleMechanics2003b, zhaoEMGdrivenMusculoskeletalModel2023a} to generate muscle forces, which are then mapped to joint torques via moment arms and integrated within a subject-specific joint dynamics model to predict multi-\gls{DoF} joint angles.

The activation estimator and the \gls{MSK} forward dynamics module are jointly optimized through end-to-end backpropagation with a composite physics-physiology loss that combines kinematic supervision, synergy regularization on measured muscles, and anatomy-guided trend regularization on unmeasured muscles.

\subsection{\gls{MSK}-NN Architecture}

\subsubsection{Muscle Activation Estimator}
The Muscle Activation Estimator employs a lightweight \gls{CNN}-based encoder.
At time step $t$, a sliding window of $L$ consecutive samples from $n_m$ \gls{sEMG} channels is first preprocessed to calculate the \gls{sEMG} envelope $u_t = [u_{i,t}]_{i=1}^{n_m}$ as described in \cite{zhaoAdaptiveCooperativeControl2023c}. The encoder takes $u_t$ as input and outputs the estimated activations for both the $n_m$ measured and $n_u$ unmeasured muscles, $\hat{a}_t = [\hat{a}_{i,t}]_{i=1}^{n_m+n_u}$.
The encoder consists of two convolutional blocks, implemented as one-dimensional convolutional layers with 32 and 64 filters (kernel size 5, stride 1, padding 2), each followed by ReLU activation and dropout. 
Global average pooling then compresses the temporal dimension into a single feature per channel, yielding a 64-dimensional feature vector. 
This vector is processed by a residual fully connected block with two layers of 128 neurons, where the output of the first hidden layer is added to the second. 
A final linear projection followed by a sigmoid activation produces instantaneous muscle activation estimates $\hat{a}_{i,t} \in [0,1]$, $\forall i\in \mathcal{M}$.

\subsubsection{Embedded \gls{MSK} Forward Dynamics Module}
\paragraph{Hill-based Muscle-tendon Dynamics}
For  $i$th muscle, the muscle–tendon force at time $t$ is given by
\begin{align}
F^{mt}_{i,t} = \big(F^a_{i,t}+F^p_{i,t}\big)\cos\varphi_{i,t}, ~~ \forall i \in \mathcal{M}
\end{align}
where $F^a_{i,t}$ and $F^p_{i,t}$ denote the active contraction force and passive force generated by muscle stretch, respectively. $\varphi_{i,t}$ is the pennation angle between the muscle fiber and the tendon. 
The active force is given by
\begin{align}
F^a_{i,t} = F_{o,i}^m \, f^a\!\left(\bar{l}_{i,t}^a\right) f^v\!\left(\bar{v}_{i,t}\right) a_{i,t}
\end{align}
where $F_{o,i}^m$ is the maximum isometric force. $f^a(\cdot)$ is the active force–length characteristic curve, 
$f^v(\cdot)$ is the force–velocity characteristic curve, and $a_{i,t}$ is the muscle activation at time $t$.
The active force–length characteristic is given by
\begin{align}
f^a(\bar{l}_{i,t}^a)=e^{-(\bar{l}_{i,t}^a-1)^2/k_a}
\end{align}
where $k_a$ is fixed at $0.45$ \cite{thelenAdjustmentMuscleMechanics2003b}, 
and normalized muscle fiber length corresponding to the activation level is given by
\begin{align}
\bar{l}^{a}_{i,t} = \frac{l^{m}_{i,t}}{l^{m}_{o,i}\left(\lambda(1-a_{i,t})+1\right)}
\end{align}
where $l^m_{i,t}$ is the current muscle fiber length, and $l^m_{o,i}$ is the optimal muscle fiber length. $\lambda$ is set to $0.15$ following \cite{lloydEMGdrivenMusculoskeletalModel2003b}.
The force–velocity characteristic is given by
\begin{equation}
 f^v(\bar{v}_{i,t}) =\left\{
\begin{aligned}
&\dfrac{0.3(\bar{v}_{i,t}+1)}{-\bar{v}_{i,t}+0.3} & \bar{v}_{i,t} \le 0 \\
&\dfrac{2.34\bar{v}_{i,t}+0.039}{1.3\bar{v}_{i,t}+0.039} & \bar{v}_{i,t} > 0
\end{aligned}
\right.
\end{equation}
where $v_{o,i}$ is the maximum shortening velocity, defined as $v_{o,i}=10\,l^m_{o,i}/\mathrm{s}$ \cite{zajacMuscleTendonProperties}. $v_{i,t}$ is the time derivative of the muscle fiber length, and $\bar{v}_{i,t}=v_{i,t}/v_{o,i}$ is the normalized fiber velocity. 
The passive muscle-stretch force is given by
\begin{align}
F^p_{i,t} =
\begin{cases}
0 & l^m_{i,t} \le l^m_{o,i}\\[6pt]
f^p\!\left(\bar{l}^m_{i,t}\right) F_{o,i}^m& l^m_{i,t} > l^m_{o,i}
\end{cases}
\end{align}
where $\bar{l}^m_{i,t} = l^m_{i,t} / l^m_{o,i}$ is the normalized muscle fiber length used for the passive force–length characteristic, 
and the passive force–length characteristic is 
\begin{align}
f^{p}\left(\bar{l}_{i,t}^{m}\right)=\frac{e^{10\left(\bar{l}_{i,t}^{m}-1\right)}}{e^{5}}
\end{align}
The pennation angle is given by
\begin{align}
\varphi_{i,t} = \arcsin\!\left(\frac{l^m_{o,i}\sin\varphi_{o,i}}{\,l^m_{i,t}\,}\right)
\end{align}
where $\varphi_{o,i}$ is the optimal pennation angle corresponding to the optimal fiber length $l^m_{o,i}$, 
and the current muscle-fiber length is given by
\begin{align}
l^m_{i,t} = \sqrt{\big(l^m_{o,i}\sin\varphi_{o,i}\big)^2 + \big(k^{mt}_i\,l^{mt}_{i,t} - l^t_{i}\big)^2}
\end{align}
where $k^{mt}_i$ is a length-scaling coefficient, and $l^{mt}_{i,t}$ is the muscle–tendon unit length approximated by polynomial fitting to OpenSim simulations \cite{delpOpenSimOpenSourceSoftware2007a}. $l^t_i$ is the tendon length, which is modeled as rigid \cite{zhaoEMGDrivenMusculoskeletalModel2020c}.

Note that, for each muscle $i \in \mathcal{M}$,  the \gls{MSK} parameters of Hill-based muscle-tendon dynamics $p_i = [F_{o,i}^m,\, l^m_{o,i},\, l^t_i,\, \varphi_{o,i},\, k^{mt}_i]^\top \in \mathbb{R}^5$ are optimized  during training through end-to-end backpropagation.

\paragraph{Joint Dynamics}
The joint dynamics are described in the general form
\begin{align}
{\hat{\tau}_t} = {M}(\hat\theta_t)\,\ddot{\hat{\theta}}_t
+ {C}(\hat\theta_t,\dot{\hat{\theta}}_t)
+ {G}(\hat\theta_t)
\label{eq:joint_dynamics}
\end{align}
where $\hat\tau_t \in \mathbb{R}^{d}$ denotes the estimated joint torque at time $t$, generated by $n$ muscles.
The estimated joint angle is $\hat\theta_t \in \mathbb{R}^d$, and $\dot{\hat{\theta}}_t, \ddot{\hat{\theta}}_t \in \mathbb{R}^{d}$ are the corresponding angular velocity and acceleration vectors obtained by numerical differentiation.
$M(\hat\theta_t) \in \mathbb{R}^{d\times d}$ is the inertia matrix, $C(\hat\theta_t,\dot{\hat\theta}_t) \in \mathbb{R}^{d}$  is the  
Coriolis and centrifugal torque,
and $G(\hat\theta_t) \in \mathbb{R}^d$ is the gravitational torque.

The joint torque is equivalently expressed as the sum of muscle contributions
\begin{equation}
{\hat{\tau}_t} = \sum_{i \in \mathcal{M}} {r}_{i}(\hat\theta_t)\, F^{mt}_{i,t}
\label{eq:muscle_torque}
\end{equation}
where ${r}_i(\hat\theta_t) \in \mathbb{R}^d$ denotes the moment arm vector of muscle $i$, which is obtained by polynomial fitting to simulation data exported from OpenSim \cite{delpOpenSimOpenSourceSoftware2007a}. The joint accelerations $\ddot{\hat{\theta}}_t$ obtained by solving \eqref{eq:joint_dynamics} and \eqref{eq:muscle_torque} are numerically integrated using a semi-implicit Euler method.

\subsection{Loss function design}
The overall loss consists of three components: a joint kinematics loss $\mathcal{L}_{\text{traj}}$, a synergy regularization loss on the measured muscles $\mathcal{L}_{\text{syn}}$, and a trend consistency loss that links unmeasured muscles with functionally related measured muscles $\mathcal{L}_{\text{trd}}$.
The joint trajectory loss penalizes discrepancies between predicted and reference angles and angular velocities, and is defined as
\begin{align}
\mathcal{L}_{\text{traj}}
= \frac{1}{T}\sum_{t=1}^{T} \Big(
\alpha\| \hat{\theta}_{t} - {\theta}_{t} \|_2^2
+ \beta\| \dot{\hat{\theta}}_{t} - \dot{\theta}_{t} \|_2^2
\Big)
\end{align}
where $T$ is the number of samples. ${\theta}_{t}$ denotes the ground truth joint angle at time $t$, while $\dot{\theta}_{t}$  is the corresponding angular velocity. $\alpha$ and $\beta$ are weighting coefficients for the position and velocity errors, respectively.

The \gls{NMF}-based synergy loss is defined as
\begin{equation}
\mathcal{L}_{\text{syn}}
=
\gamma\frac{1}{T}
\sum_{t=1}^{T}
\min_{{s}_t \ge 0}
\|  \hat{a}^{\mathcal{M}_m}_t - {W}{s}_t \|_2^2
\end{equation}
where $\hat{ a}^{\mathcal{M}_m}_t = [\hat{a}_{i,t}]_{i=1}^{n_m} \in \mathbb{R}^{n_m}$ denotes the predicted activations of the measured muscles at time $t$.
${W} \in \mathbb{R}^{n_m \times r}$ is the synergy basis extracted offline via \gls{NMF} from the recorded \gls{sEMG} envelopes of the measured muscles in the training set, where $r$ denotes the dimensionality of the retained synergy subspace.
${W}$ is kept frozen during network optimization to serve as 
a fixed physiological reference, and ${s}_t \in \mathbb{R}^{r}_{\ge 0}$ denotes the corresponding non-negative synergy coefficients obtained from the inner minimization. Intuitively, $\mathcal{L}_{\text{syn}}$ penalizes the distance from $\hat{a}^{\mathcal{M}_m}_t$ to the non-negative cone spanned by ${W}$, thereby constraining the predicted activations 
to lie within the physiologically plausible coordination subspace learned from \gls{sEMG} data.
Since ${W}$ is fixed, the inner minimization is solved as a non-negative least-squares problem at each time step, and the resulting synergy loss remains differentiable with respect to $\hat{a}^{\mathcal{M}_m}_t$. $\gamma$ is the weighting coefficient for the synergy loss term.

The trend loss couples each unmeasured muscle with anatomically related measured muscles, encouraging their activation patterns to share temporal coordination structure.
For each unmeasured muscle $i \in \mathcal{M}_u$, define its anatomically related measured muscles as $\mathcal{R}_i \subseteq \mathcal{M}_m$, then the complete set of paired muscles can be defined as $\mathcal{P} = \{(i,j) \mid i \in \mathcal{M}_u, \, j \in \mathcal{R}_i\}$.  
To capture coordinated temporal dynamics rather than absolute amplitudes, we define the activation temporal differences as $\Delta \hat{a}_{i,t} = \hat{a}_{i,t} - \hat{a}_{i,t-1}$ for $t = 2, \ldots, T$ and $i \in \mathcal{M}$.

The trend loss is then formulated as
\begin{align}
\mathcal{L}_{\mathrm{trd}} 
&=  \frac{\delta}{|\mathcal{P}|}\sum_{(i,j)\in\mathcal{P}} \big(1-\rho_{i,j}\big) \\
\rho_{i,j} &=
\frac{\sum_{t=2}^{T} \Delta \hat{a}_{i,t} \, \Delta \hat{a}_{j,t}}
{\sqrt{\sum_{t=2}^{T} (\Delta \hat{a}_{i,t})^2} \;
 \sqrt{\sum_{t=2}^{T} (\Delta \hat{a}_{j,t})^2}}
\end{align}
where $\rho_{i,j} \in [-1, 1]$ measures the temporal correlation between the activation differences of muscles $i$ and $j$, and $\delta$ is the weighting coefficient for the trend consistency loss.
The loss is minimized when paired unmeasured and measured muscles exhibit aligned moment-to-moment activation changes, transferring temporal coordination structure from observed to unobserved muscles.

\section{Experimental Setup and Evaluation Protocol}

\subsection{Experimental Scenario and Muscle Configuration}
The proposed framework is validated on four wrist motions involving both \gls{FE} and \gls{RUD} at unconstrained speeds and amplitudes: \gls{CW}; \gls{CCW}; $\infty$-shaped motion; \gls{RND}.

Six muscle-tendon units are considered in the embedded \gls{MSK} forward dynamics module: \gls{FCR}, \gls{FCU}, \gls{ECRL}, \gls{ECRB}, \gls{ECU}, \gls{APL}. For notation consistency, muscle indices are assigned as follows: \gls{FCR} ($i=1$), \gls{FCU} ($i=2$), \gls{ECRL} ($i=3$), \gls{ECRB} ($i=4$), \gls{ECU} ($i=5$), and \gls{APL} ($i=6$).
The first five are superficial muscles commonly adopted in \gls{sEMG}-based wrist \gls{MSK} modeling \cite{zhaoEMGDrivenMusculoskeletalModel2020c}. 
\gls{APL} is a deep muscle that cannot be measured by \gls{sEMG} \cite{eschweilerAnatomyBiomechanicsLoads2022a}, yet it plays a critical role in radial deviation torque generation \cite{shahImportanceAbductorPollicis2018a}. It is therefore included in the model as an anatomically unmeasured muscle. In addition, although \gls{ECRB} can be measured by \gls{sEMG}, its signal is intentionally excluded from the input of \gls{MSK}-NN to define an artificially unmeasured muscle. The recorded \gls{sEMG} envelope of \gls{ECRB} is then used as a ground-truth reference to validate the physiological plausibility of the corresponding activation estimated by \gls{MSK}-NN. 
For the unmeasured muscles $i \in \{4,6\}$ (i.e., \gls{ECRB} and \gls{APL}), the anatomically related measured-muscle sets are defined by index as $\mathcal{R}_{4}=\{3\}$ and $\mathcal{R}_{6}=\{1\}$, reflecting their functional grouping based on moment-arm contributions to the corresponding \gls{DoF}.

\begin{table}[!t]
\centering
\caption{Initial value of $p_i$ for six muscles}
\label{tab:hill_parameters}

\renewcommand{\arraystretch}{1.0}
\begin{tabular}{c c c c c c}
\toprule
Muscle & $l^m_{o,i}$ (m) & $F^m_{o,i}$ (N) & $l^t_i$ (m) & $\varphi_{o,i}$ (rad) & $k^{mt}_i$ \\
\midrule
FCR  & 0.0628 & 59.7  & 0.2185 & 0.054 & 1 \\
FCU  & 0.0510 & 102.6 & 0.2420 & 0.210 & 1 \\
ECRL & 0.0936 & 65.7  & 0.2026 & 0.044 & 1 \\
ECRB & 0.0585 & 122.9 & 0.2205 & 0.155 & 1 \\
ECU  & 0.0620 & 117.0   & 0.2310 & 0.061 & 1 \\
APL  & 0.0710 & 57.9  & 0.1250 & 0.130 & 1 \\
\bottomrule
\end{tabular}
\end{table}

\subsection{\gls{MSK} Parameter Initialization}
The \gls{MSK} parameters $p_i$, for $i=1,\ldots,6$, are initialized based on the OpenSim hand-wrist model~\cite{gonzalezHowMuscleArchitecture1997}, as summarized in Table~\ref{tab:hill_parameters}. The parameters are further optimized during training through end-to-end backpropagation, allowing subject-specific personalization of muscle-tendon dynamics.

For the two-\gls{DoF} wrist implementation, the Coriolis and centrifugal terms are omitted given their negligible contribution to wrist rotations~\cite{charlesDynamicsWristRotations2011}. 
The remaining velocity-dependent joint resistance is approximated as a linear viscous damping term, with coefficients adopted from in vivo wrist dynamics measurements~\cite{parkVivoEstimationHuman2017}.
The inertia matrix and gravitational torque are computed from subject-specific anthropometric measurements, including body mass and hand length, following established segment inertial parameter models \cite{delevaAdjustmentsZatsiorskySeluyanovsSegment1996}. 

Following \gls{NMF}-based muscle synergy analysis~\cite{treschMatrixFactorizationAlgorithms2006}, the synergy dimension $r$ of ${W}$ in $\mathcal{L}_{\mathrm{syn}}$ is determined as the minimum value achieving a \gls{VAF} of at least 0.90 in reconstructing the measured-muscle \gls{sEMG} envelopes, yielding $r = 3$ across all subjects in this study.

\subsection{Data Acquisition and Processing}
The experimental protocol is approved by the University Research Ethics Committee at the University of Manchester (Reference No. 2024-20628-36971).
Six healthy subjects participated in the study with written informed consent. 
Height, body mass, and hand length are recorded before data collection.
During the experiment, participants perform wrist movements in a seated position while maintaining a neutral forearm posture and avoiding forearm pronation and supination. 

\begin{figure}[!t]
    \centering
    \includegraphics[width=0.44\textwidth]{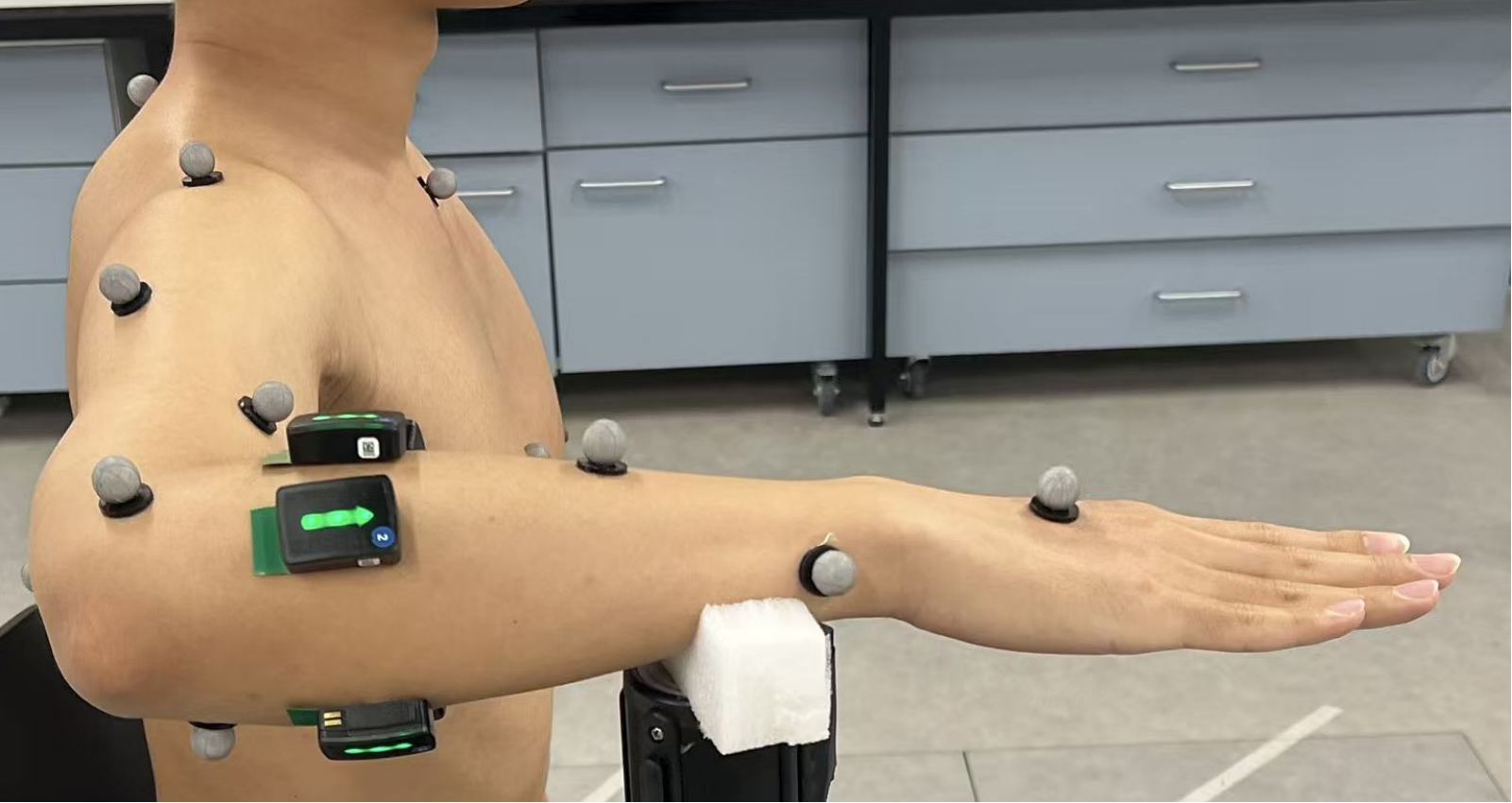}
    \caption{Experiment setup: $14$ reflective markers are attached on subject’s right upper limb. Electrodes are placed on five primary muscles of wrist joint, including \gls{FCR}, \gls{FCU}, \gls{ECRL}, \gls{ECRB} and \gls{ECU}.}
    \label{fig:Experiment}
\end{figure}

Kinematic data are captured using a Vicon motion capture system (Vicon Motion Systems Ltd., Oxford, UK) at $250$ Hz. 14 reflective markers are placed according to the Vicon upper-limb model protocol, as illustrated in Fig.~\ref{fig:Experiment}. 
\gls{sEMG} signals are recorded from five wrist muscles (\gls{FCR}, \gls{FCU}, \gls{ECRL}, \gls{ECRB}, and \gls{ECU}) using Delsys Avanti sensors (Delsys Inc., Natick, MA, USA) at $2000$ Hz. Electrode placement follows the SENIAM recommendations \cite{stegemanStandardsSurfaceElectromyography}. The two systems are hardware-synchronized via a trigger interface.

Each participant performs four types of continuous wrist motion patterns without constraints on speed or amplitude: \gls{CW}, \gls{CCW}, $\infty$-shaped, and \gls{RND}.
For each pattern, six $40$ s trials are recorded, with the middle $30$s retained to remove transient effects.

Raw \gls{sEMG} signals are band-pass filtered (20--450 Hz, fourth-order Butterworth), rectified, and low-pass filtered (4 Hz, fourth-order Butterworth) to obtain \gls{sEMG} envelopes. 
\Gls{MVC} of each muscle is recorded before the experiment, and the envelopes are normalized by the \gls{MVC} peak amplitude of each muscle. Processed \gls{sEMG} envelopes are resampled to $250$ Hz to match the kinematic sampling rate.

\subsection{MSK-NN Training}
The preprocessed dataset is split at the trial level into training, validation, and test sets with a 4:1:1 ratio.
A subject-specific model is trained using all four motions pooled together, with evaluation conducted separately for each motion type.

During training, sequences from the training set are divided into overlapping segments of 64 time steps with a step size of 32.
Each \gls{CNN} input consists of a sliding \gls{sEMG} window of $L = 32$ time steps, advanced with a stride of 1.
This segmentation increases the number of training samples and stabilizes optimization by averaging the loss over time within each segment.
Validation and testing are performed on full sequences using continuous sequential inference without segment-level resetting.
The model is optimized using AdamW with a learning rate of $1\times10^{-4}$ \cite{loshchilov2019decoupled}.
The loss weighting coefficients are set to $\alpha=0.6$, $\beta=0.2$, $\gamma=0.1$, and $\delta=0.1$.
A batch size of 64 is used, and a dropout rate of 0.2 is applied to the neural encoder for regularization.
Early stopping with a patience of 30 epochs is applied based on validation performance.

\subsection{Evaluation Criteria}
For each \gls{DoF}, kinematics estimation performance is evaluated using \gls{NRMSE} for amplitude discrepancy and $R^2$ for correlation with ground truth.
\begin{align}
\mathrm{NRMSE}&=\frac{\sqrt{\tfrac{1}{T}\sum_{t=1}^T (\theta_t - {\hat{\theta}}_t)^2}}
     {\theta_{\max} - \theta_{\min}}\\
R^2&=1-\frac{\sum_{t=1}^T(\theta_t-{\hat{\theta}}_t)^2}{\sum_{t=1}^T(\theta_t-\bar{\theta})^2}
\end{align}
where $\theta_{\max} = \max_{1 \leq t \leq T} \theta_t$ and $\theta_{\min} = \min_{1 \leq t \leq T} \theta_t$.

Physiological plausibility of the estimated muscle activation is evaluated using \gls{CC} and \gls{SC} between the estimated activation of \gls{ECRB} and the measured \gls{sEMG} envelope of \gls{ECRB}. For $i$th muscle, the \gls{CC} and \gls{SC} are defined as
\begin{align}
\mathrm{CC}_i &= \frac{\sum_{t=1}^T \big(a_{i,t} - \bar{a}_i\big)\big(u_{i,t} - \bar{u}_i\big)}
{\sqrt{\sum_{t=1}^T \big(a_{i,t} - \bar{a}_i\big)^2 \sum_{t=1}^T \big(u_{i,t} - \bar{u}_i\big)^2}}\\
\mathrm{SC}_i &= \frac{\sum_{t=1}^T \big(\mathcal{T}(a_{i,t}) - \bar{\mathcal{T}}_{a,i}\big)\big(\mathcal{T}(u_{i,t}) - \bar{\mathcal{T}}_{u,i}\big)}
{\sqrt{\sum_{t=1}^T \big(\mathcal{T}(a_{i,t}) - \bar{\mathcal{T}}_{a,i}\big)^2 \sum_{t=1}^T \big(\mathcal{T}(u_{i,t}) - \bar{\mathcal{T}}_{u,i}\big)^2}}
\end{align}
where $T$ is the number of samples, and $\bar{a}_i$ and $\bar{u}_i$ denote the mean values of the estimated activation and recorded \gls{sEMG} envelope for the $i$th muscle, respectively. $\mathcal{T}(\cdot)$ denotes rank transformation, which maps each element of a sequence to its rank in ascending order. \(\bar{\mathcal{T}}_{a,i}\) and \(\bar{\mathcal{T}}_{u,i}\) denote the mean ranks of
\(\{\mathcal{T}(a_{i,t})\}_{t=1}^{T}\) and
\(\{\mathcal{T}(u_{i,t})\}_{t=1}^{T}\), respectively.

\subsection{Baseline Methods}
The proposed \gls{MSK}-NN is compared with four data-driven baselines: a \gls{CNN} regressor (comparison for showing the contribution of embedded \gls{MSK} dynamics), \gls{Bi-LSTM} \cite{maBiDirectionalLSTMNetwork2021b} (recurrent temporal-modeling baseline), \gls{CNN}-LSTM \cite{baoCNNLSTMHybridModel2021b} (convolutional--recurrent spatiotemporal baseline), and \gls{PET} \cite{linParallelEfficientTransformer2025} (transformer-based sequence-modeling baseline).
All baselines are implemented and trained on the same dataset using the same subject-wise data splits, input \gls{sEMG} channels, and two-\gls{DoF} wrist-angle targets as \gls{MSK}-NN.
Baseline hyperparameters follow the cited implementations where applicable and adjusted only to match the input dimension and prediction horizon of the present task.
\begin{figure*}[!t]
    \centering
    \includegraphics[width=0.75\textwidth]{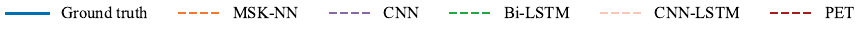}

    \vspace{0.4em}    
     \begin{subfigure}{\textwidth}
        \centering
        \includegraphics[width=\textwidth]{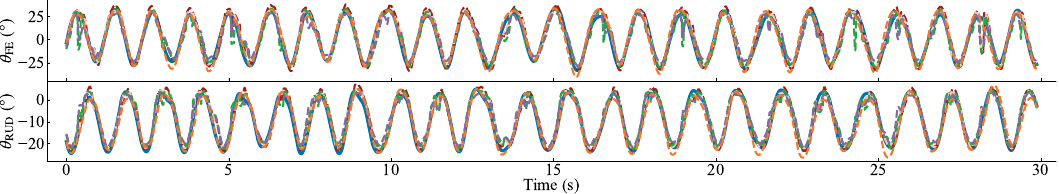}
        \caption{\gls{CW}}
        \label{fig:comparison_cw_ccw}
    \end{subfigure}

     \begin{subfigure}{\textwidth}
        \centering
        \includegraphics[width=\textwidth]{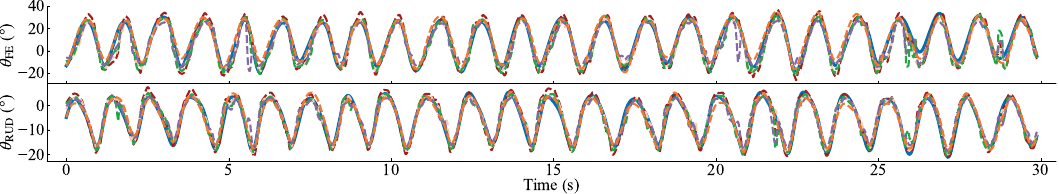}
        \caption{\gls{CCW}}
        \label{fig:comparison_cw_ccw}
    \end{subfigure}
    
    \begin{subfigure}{\textwidth}
        \centering
        \includegraphics[width=\textwidth]{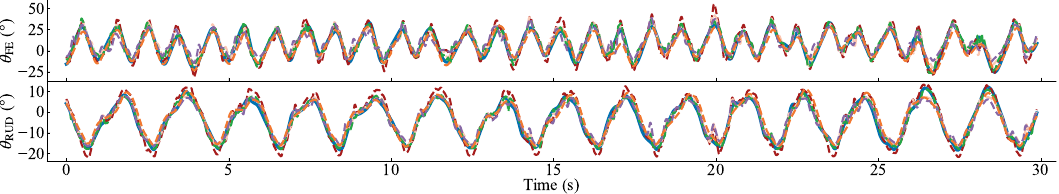}
        \caption{$\infty$-shaped motion}
        \label{fig:comparison_cw_ccw}
    \end{subfigure}
    
    \begin{subfigure}{\textwidth}
        \centering
        \includegraphics[width=\textwidth]{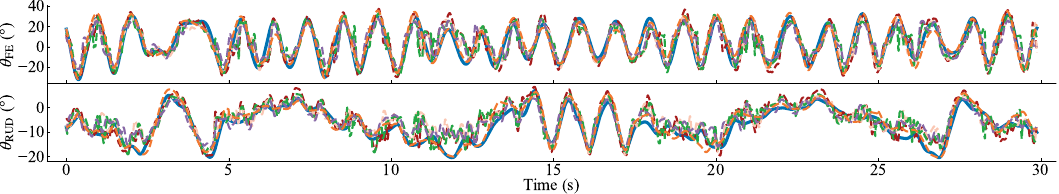}
        \caption{\gls{RND} motion}
        \label{fig:comparison_cw_ccw}
    \end{subfigure}

    \caption{
    Comparison of estimated wrist joint angles by \gls{MSK}-NN, \gls{CNN},
    \gls{Bi-LSTM}, \gls{CNN}-LSTM, and \gls{PET} against the ground truth
    for a representative subject (S1) across four motions.
    }
    \label{fig:comparison}
\end{figure*}
\begin{table*}[t]
\centering
\caption{NRMSE comparison across four wrist motions.}
\label{tab:nrmse_models}
\setlength{\tabcolsep}{4pt}

\begin{tabular}{c *{8}{c}}
\toprule
\multirow{2}{*}{\textbf{Methods}}
& \multicolumn{2}{c}{\textbf{CW}}
& \multicolumn{2}{c}{\textbf{CCW}}
& \multicolumn{2}{c}{\textbf{$\infty$-shaped}}
& \multicolumn{2}{c}{\textbf{\gls{RND}}} \\

\cmidrule(lr){2-3} \cmidrule(lr){4-5} \cmidrule(lr){6-7} \cmidrule(lr){8-9}
& FE & RUD & FE & RUD & FE & RUD & FE & RUD \\
\midrule

CNN
& 0.123 $\pm$ 0.026 & 0.128 $\pm$ 0.022
& 0.131 $\pm$ 0.019 & 0.116 $\pm$ 0.024
& 0.116 $\pm$ 0.011 & 0.109 $\pm$ 0.024
& 0.143 $\pm$ 0.026 & 0.143 $\pm$ 0.023 \\

Bi-LSTM
& 0.123 $\pm$ 0.024 & 0.127 $\pm$ 0.014
& 0.121 $\pm$ 0.015 & 0.113 $\pm$ 0.011
& 0.101 $\pm$ 0.016 & 0.109 $\pm$ 0.018
& 0.153 $\pm$ 0.030 & 0.143 $\pm$ 0.015 \\

CNN-LSTM
& 0.109 $\pm$ 0.021 & 0.107 $\pm$ 0.014
& 0.094 $\pm$ 0.015 & 0.095 $\pm$ 0.018
& 0.101 $\pm$ 0.024 & 0.092 $\pm$ 0.017
& 0.119 $\pm$ 0.026 & 0.125 $\pm$ 0.009 \\

PET
& 0.101 $\pm$ 0.018 & 0.101 $\pm$ 0.019
& 0.110 $\pm$ 0.031 & 0.109 $\pm$ 0.037
& 0.091 $\pm$ 0.016 & 0.089 $\pm$ 0.016
& 0.134 $\pm$ 0.032 & 0.121 $\pm$ 0.019 \\

MSK-NN
& \textbf{0.079 $\pm$ 0.013} & \textbf{0.088 $\pm$ 0.014}
& \textbf{0.080 $\pm$ 0.010} & \textbf{0.083 $\pm$ 0.018}
& \textbf{0.078 $\pm$ 0.012} & \textbf{0.083 $\pm$ 0.010}
& \textbf{0.091 $\pm$ 0.020} & \textbf{0.101 $\pm$ 0.009} \\

\bottomrule
\end{tabular}
\end{table*}

\begin{table*}[t]
\centering
\caption{$R^2$ comparison across four wrist motions.}
\label{tab:r2_models}

\setlength{\tabcolsep}{4pt}

\begin{tabular}{c *{8}{c}}
\toprule
\multirow{2}{*}{\textbf{Methods}}
& \multicolumn{2}{c}{\textbf{CW}}
& \multicolumn{2}{c}{\textbf{CCW}}
& \multicolumn{2}{c}{\textbf{$\infty$-shaped}}
& \multicolumn{2}{c}{\textbf{\gls{RND}}} \\

\cmidrule(lr){2-3} \cmidrule(lr){4-5} \cmidrule(lr){6-7} \cmidrule(lr){8-9}
& FE & RUD & FE & RUD & FE & RUD & FE & RUD \\

\midrule

CNN
& 0.783 $\pm$ 0.087 & 0.807 $\pm$ 0.079
& 0.763 $\pm$ 0.083 & 0.813 $\pm$ 0.079
& 0.770 $\pm$ 0.062 & 0.818 $\pm$ 0.064
& 0.645 $\pm$ 0.110 & 0.627 $\pm$ 0.115 \\

Bi-LSTM
& 0.793 $\pm$ 0.068 & 0.808 $\pm$ 0.051
& 0.800 $\pm$ 0.035 & 0.805 $\pm$ 0.019
& 0.823 $\pm$ 0.066 & 0.822 $\pm$ 0.061
& 0.575 $\pm$ 0.170 & 0.587 $\pm$ 0.092 \\

CNN-LSTM
& 0.830 $\pm$ 0.075 & 0.860 $\pm$ 0.040
& 0.880 $\pm$ 0.042 & 0.878 $\pm$ 0.043
& 0.838 $\pm$ 0.060 & 0.872 $\pm$ 0.050
& 0.723 $\pm$ 0.120 & 0.687 $\pm$ 0.060 \\

PET
& 0.867 $\pm$ 0.046 & 0.885 $\pm$ 0.043
& 0.825 $\pm$ 0.133 & 0.810 $\pm$ 0.115
& 0.862 $\pm$ 0.050 & 0.883 $\pm$ 0.034
& 0.665 $\pm$ 0.174 & 0.710 $\pm$ 0.092 \\

MSK-NN
& \textbf{0.918 $\pm$ 0.029} & \textbf{0.910 $\pm$ 0.040}
& \textbf{0.925 $\pm$ 0.019} & \textbf{0.916 $\pm$ 0.032}
& \textbf{0.900 $\pm$ 0.045} & \textbf{0.900 $\pm$ 0.040}
& \textbf{0.835 $\pm$ 0.057} & \textbf{0.803 $\pm$ 0.051} \\

\bottomrule
\end{tabular}
\end{table*}

\begin{figure*}[t]
    \centering
    \includegraphics[width=0.54\textwidth]{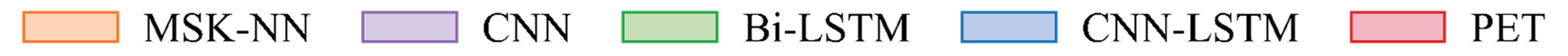}

    \vspace{0.4em}    
    \begin{subfigure}{0.48\textwidth}
        \centering
        \includegraphics[width=\linewidth]{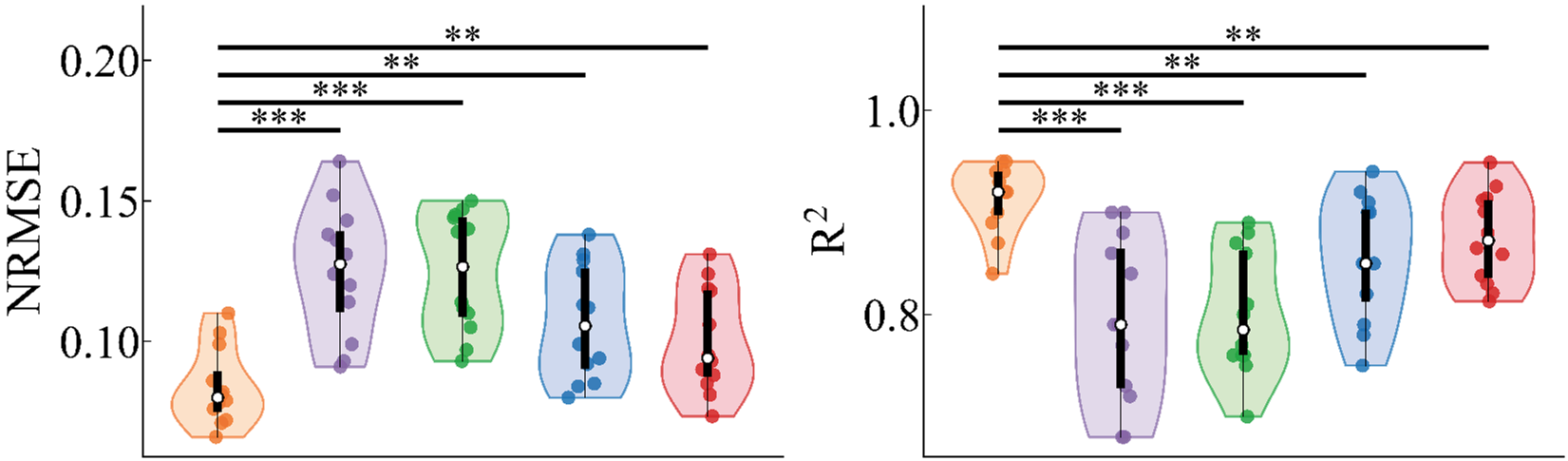}
        \caption{\gls{CW}}
        \label{fig:activation_cw}
    \end{subfigure}%
    \hfill
    \begin{subfigure}{0.48\textwidth}
        \centering
        \includegraphics[width=\linewidth]{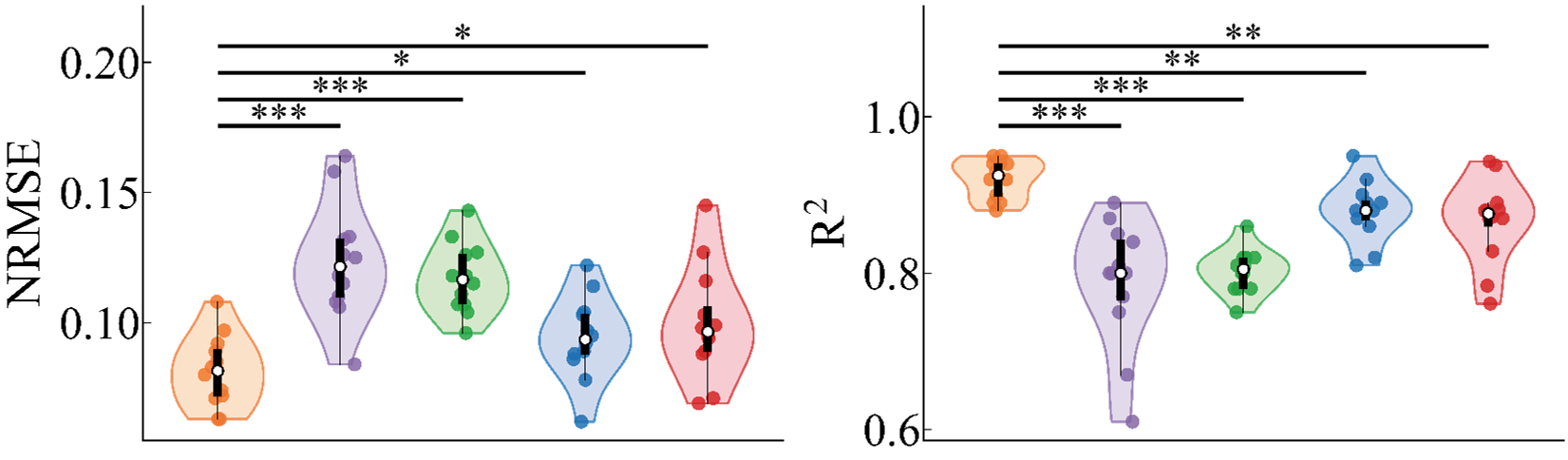}
        \caption{\gls{CCW}}
        \label{fig:activation_ccw}
    \end{subfigure}

    \vspace{0.1em}

    \begin{subfigure}{0.48\textwidth}
        \centering
        \includegraphics[width=\linewidth]{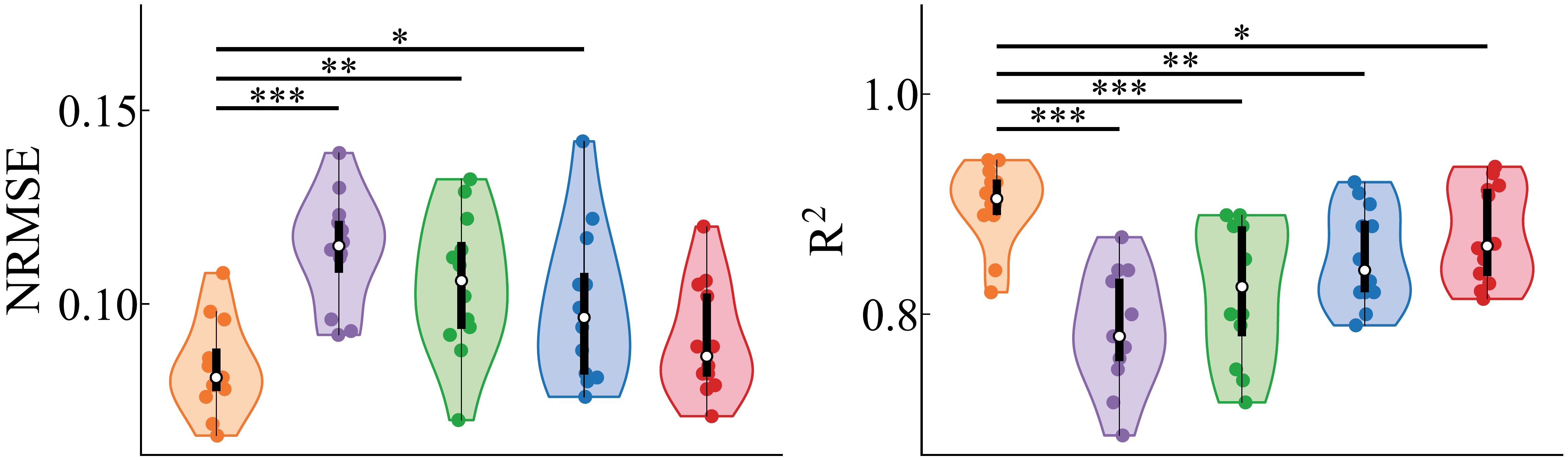}
        \caption{$\infty$-shaped motion}
        \label{fig:activation_8}
    \end{subfigure}%
    \hfill
    \begin{subfigure}{0.48\textwidth}
        \centering
        \includegraphics[width=\linewidth]{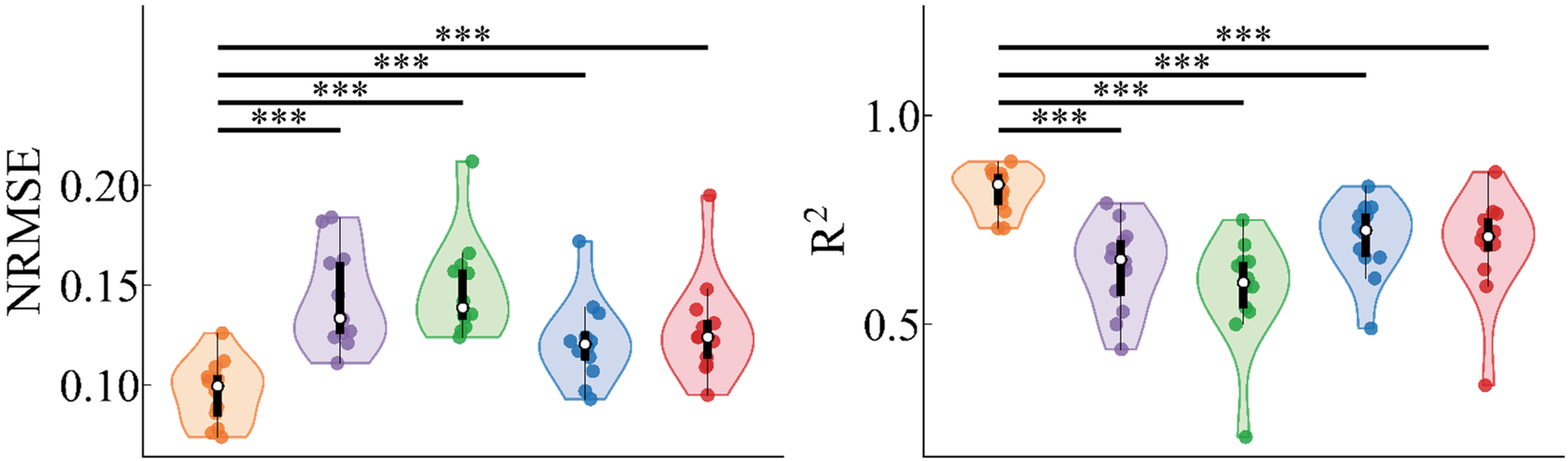}
        \caption{\gls{RND} motion}
        \label{fig:activation_random}
    \end{subfigure}

\caption{Comparison of \gls{MSK}-NN and baseline methods across four wrist motions. Violin plots illustrate the distribution of subject-wise performance over six subjects, with colored dots representing individual subjects. The thick black bar indicates the interquartile range and the white dot marks the median value. Asterisks indicate the statistically significant difference between methods (*: $p < 0.05$, **: $p < 0.01$, and ***: $p < 0.001$).
    }
    \label{fig:Paired_t}
\end{figure*}

\section{Results}
The proposed \gls{MSK}-NN is evaluated from four perspectives.
First, its multi-\gls{DoF} kinematics prediction performance is compared against state-of-the-art deep learning baselines.
Second, the inferred input-excluded \gls{ECRB} activation is compared with the recorded \gls{ECRB} \gls{sEMG} envelope, which is withheld from the input and used only for physiological plausibility assessment.
Third, the optimized \gls{MSK} parameters are examined against physiological ranges.
Finally, a muscle ablation analysis is conducted to assess the contribution of unmeasured muscles to kinematics estimation performance.

\subsection{Kinematics Decoding Performance Comparison}
Fig.~\ref{fig:comparison} compares the wrist joint angles $\theta_{\mathrm{FE}}$ and $\theta_{\mathrm{RUD}}$ estimated by \gls{MSK}-NN and the baseline methods against the ground truth across the four motions for subject S1.
\gls{MSK}-NN aligns most closely with the ground truth in both \glspl{DoF}, preserving amplitude and temporal consistency across the four motions.
For \gls{CW}, \gls{CCW}, and $\infty$-shaped motions, the baselines mainly show local deviations near peaks and rapid transitions.
This tendency becomes more pronounced under \gls{RND}, especially in segments with more irregular and rapid variations, as reflected more clearly in \gls{RUD}.
In contrast, \gls{MSK}-NN maintains smoother and more consistent tracking, demonstrating stronger prediction robustness.

Tables~\ref{tab:nrmse_models} and~\ref{tab:r2_models} summarize the \gls{NRMSE} and \(R^2\) averaged over six subjects.
\gls{MSK}-NN consistently achieves the lowest \gls{NRMSE} and highest \(R^2\) across all motions and both \glspl{DoF}, indicating superior overall estimation accuracy.
The smaller standard deviations further suggest more consistent performance across subjects.
\Gls{RND} represents the most challenging condition, as all methods show increased errors compared with the more structured motions.
Notably, under \gls{RND}, the baseline \(R^2\) values drop markedly, whereas \gls{MSK}-NN remains above $0.80$ in both \gls{FE} and \gls{RUD}.

Paired $t$-tests with Holm–Bonferroni correction are conducted between \gls{MSK}-NN and each baseline.
For each motion, \gls{NRMSE} and \gls{R2} are analyzed separately using pooled \gls{FE} and \gls{RUD} results from all subjects, with results summarized in Fig.~\ref{fig:Paired_t}.
\gls{MSK}-NN significantly outperforms \gls{CNN} and \gls{Bi-LSTM} across all conditions. 
Since the \gls{CNN} baseline isolates the activation-estimator component of \gls{MSK}-NN, the consistent performance gains of \gls{MSK}-NN over this baseline highlight the contribution of the embedded \gls{MSK} forward dynamics module.
For \gls{CNN}-LSTM and \gls{PET}, significant improvements are observed in most comparisons, except for the $\infty$-shaped \gls{RUD} comparison against \gls{PET}.
The significance levels are generally stronger under \gls{RND}, further highlighting the advantage of \gls{MSK}-NN in more challenging movement conditions.

\begin{figure}[t]
\centering

\begin{subfigure}{\columnwidth}
    \centering
    \includegraphics[width=\columnwidth]{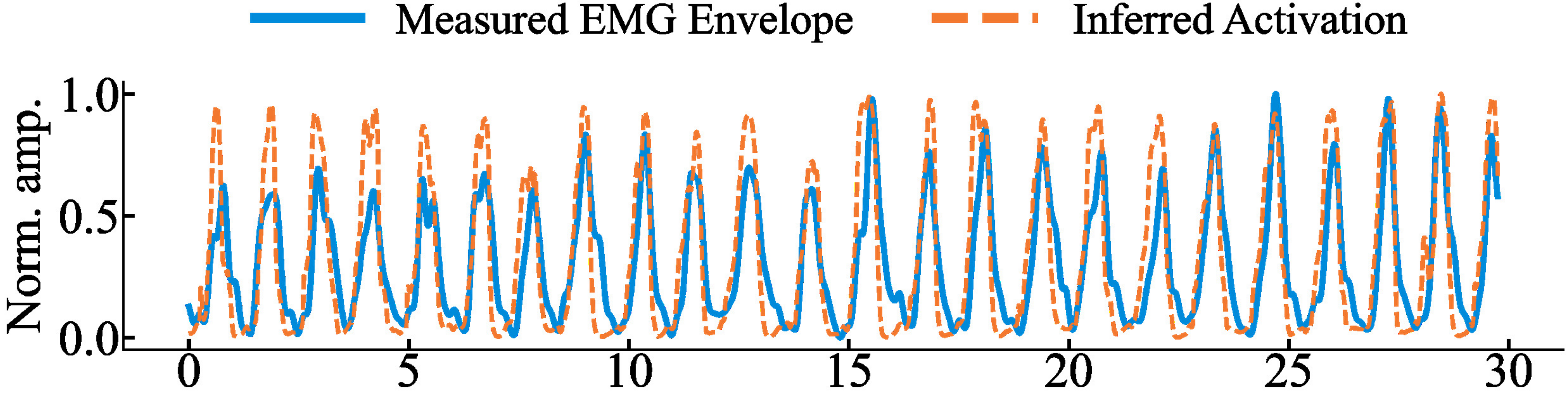}
    \caption{\gls{CW}}
    \label{fig:activation_comparison_a}
\end{subfigure}

\begin{subfigure}{\columnwidth}
    \centering
    \includegraphics[width=\columnwidth]{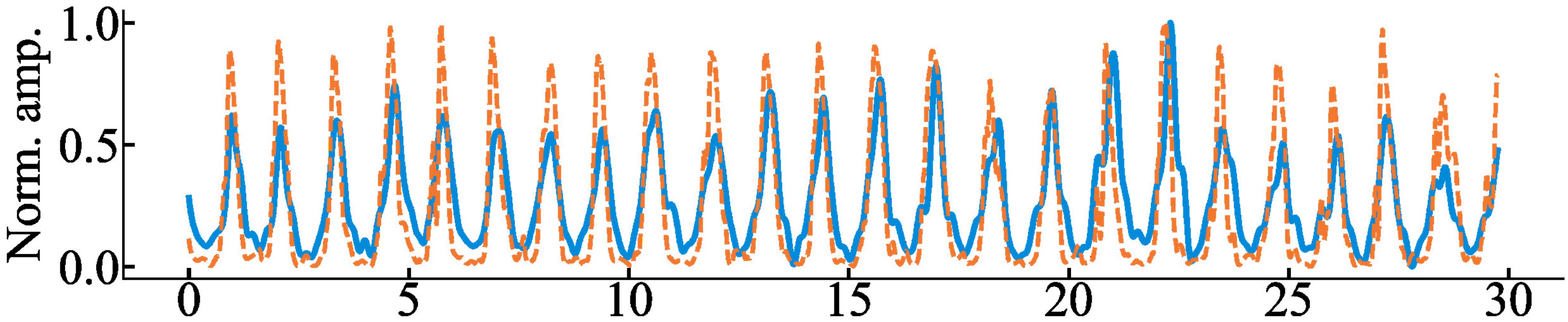}
    \caption{\gls{CCW}}
    \label{fig:activation_comparison_b}
\end{subfigure}

\begin{subfigure}{\columnwidth}
    \centering
    \includegraphics[width=\columnwidth]{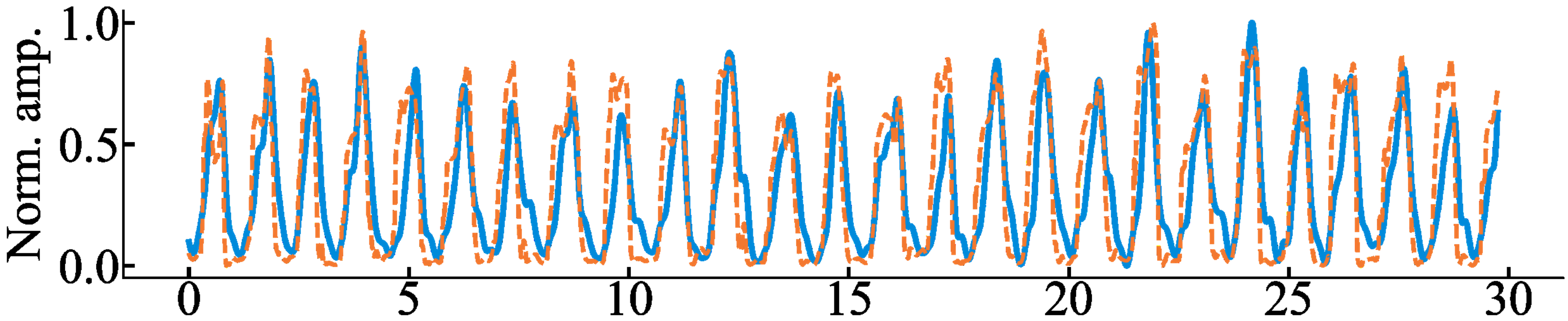}
    \caption{$\infty$-shaped motion}
    \label{fig:activation_comparison_c}
\end{subfigure}

\begin{subfigure}{\columnwidth}
    \centering
    \includegraphics[width=\columnwidth]{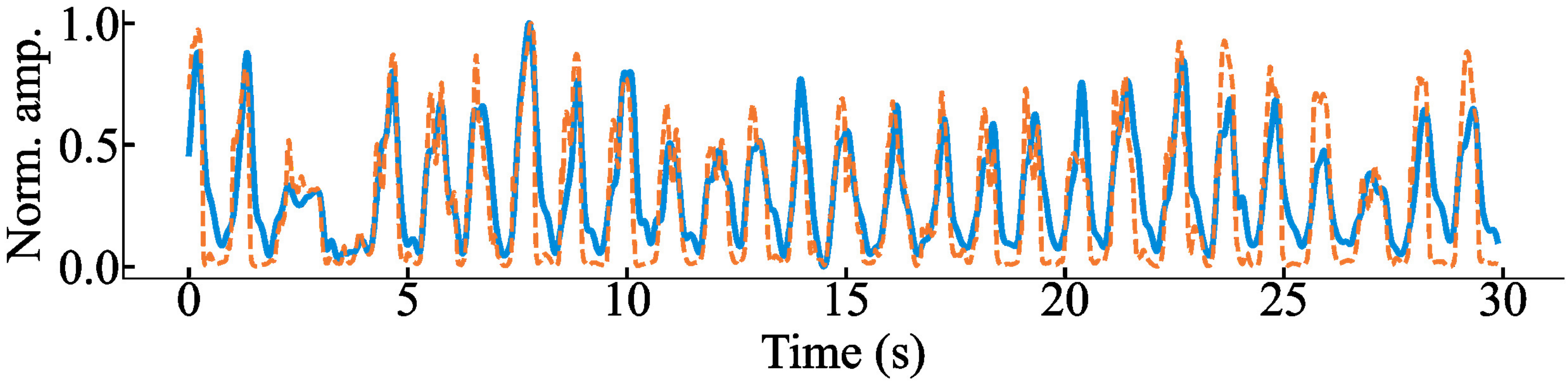}
    \caption{\gls{RND} motion}
    \label{fig:activation_comparison_d}
\end{subfigure}

\caption{
Comparison of the min--max normalized recorded \gls{sEMG} envelope and estimated activation of \gls{ECRB} for a representative subject (S1) across four motions.
}
\label{fig:activation_comparison}
\end{figure}

\subsection{Unmeasured Muscle Activation Consistency}
\label{subsec:unmeasured}
\begin{figure}[t]
    \centering
    \includegraphics[width=0.94\columnwidth,keepaspectratio]{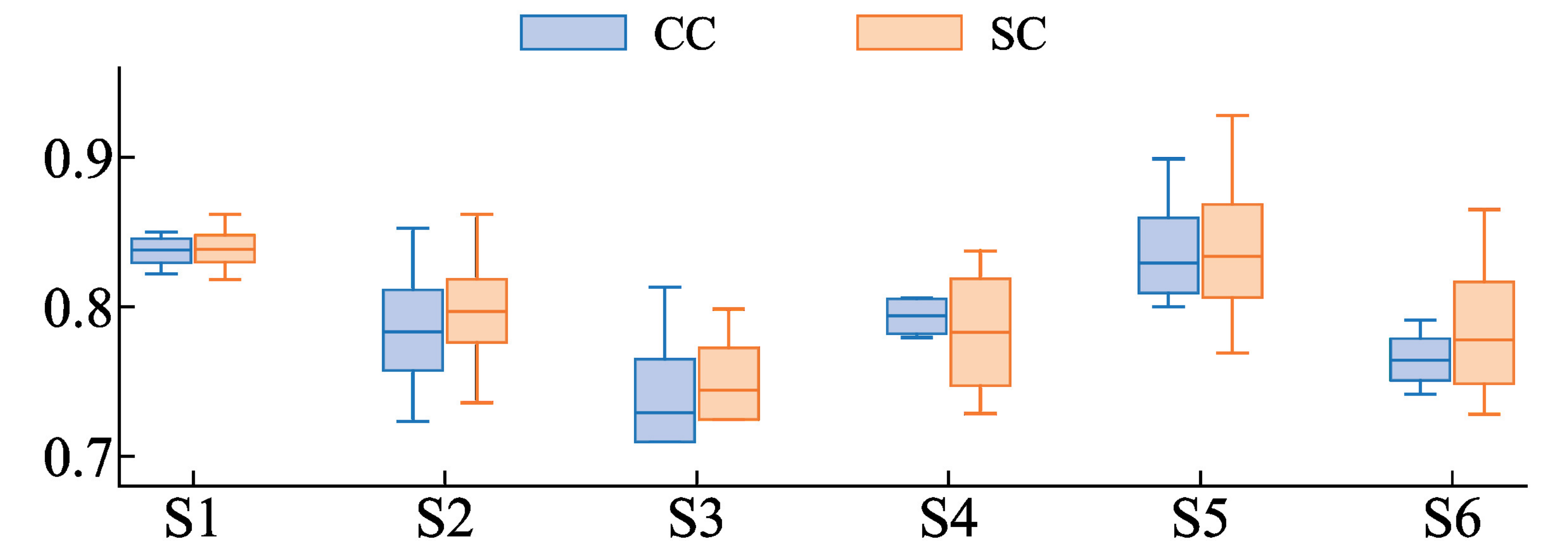}
    \caption{
    Distribution of \gls{CC} and \gls{SC} between the estimated activation and recorded \gls{sEMG} envelope of \gls{ECRB} across all subjects. Each box summarizes results aggregated over four motions. 
    }
    \label{fig:activation_emg_heatmap}
\end{figure}

Latent activation plausibility is assessed by comparing the inferred input-excluded \gls{ECRB} activation with its recorded \gls{sEMG} envelope after amplitude normalization.
Fig.~\ref{fig:activation_comparison} shows representative time-series results for Subject~1 across the four wrist motions.
The inferred activation generally follows the recorded \gls{sEMG} envelope, capturing activation timing, peak locations, and cyclic patterns across both structured and random movements.
Fig.~\ref{fig:activation_emg_heatmap} summarizes the \gls{CC} and \gls{SC} between the inferred activation and recorded \gls{sEMG} envelope across all subjects and motions.
All \gls{CC} values fall within the strong correlation range ($>0.67$) defined in \cite{aoEMGdrivenMusculoskeletalModel2022a,taylorInterpretationCorrelationCoefficient1990}.
High \gls{CC} and \gls{SC} values indicate agreement in amplitude co-variation and temporal trend consistency, with limited variability across subjects and motion types.
These results suggest that \gls{MSK}-NN can infer physiologically plausible latent activations for unmeasured muscles under partially observed \gls{sEMG}.

\subsection{\gls{MSK} Parameter Personalization}

\begin{table*}[t]
\centering
\caption{Optimized \gls{MSK} parameters of a representative subject (S3) vs physiological limits.}
\label{tab:optimized_subject4}
\renewcommand{\arraystretch}{1.0}
\begin{tabular*}{\textwidth}{@{\extracolsep{\fill}}c *{10}{c}}
\toprule
& \multicolumn{2}{c}{$l^{m}_{o,i}$ (m)} 
& \multicolumn{2}{c}{$F^{m}_{o,i}$ (N)} 
& \multicolumn{2}{c}{$l^{t}_{i}$ (m)} 
& \multicolumn{2}{c}{$\varphi_{o,i}$ (rad)} 
& \multicolumn{2}{c}{$k^{mt}_{i}$} \\
\cmidrule(lr){2-3} \cmidrule(lr){4-5} \cmidrule(lr){6-7} \cmidrule(lr){8-9} \cmidrule(lr){10-11}
\textbf{Muscle index} & Optimized & Range & Optimized & Range & Optimized & Range & Optimized & Range & Optimized & Range \\
\midrule
FCR  & 0.0624 & 0.052--0.072 & 61.0 & 30.5--91.5 & 0.2163 & 0.206--0.226 & 0.0543 & 0.051--0.056 & 0.918 & 0.9--1.2 \\
FCU  & 0.0509 & 0.041--0.061 & 101.4 & 50.7--152.1 & 0.2410 & 0.231--0.251 & 0.2105 & 0.200--0.221 & 0.937 & 0.9--1.2 \\
ECRL & 0.0940 & 0.084--0.104 & 64.8 & 32.4--97.2 & 0.2042 & 0.194--0.214 & 0.0433 & 0.041--0.046 & 0.973 & 0.9--1.2 \\
ECRB & 0.0581 & 0.048--0.068 & 118.3 & 59.2--177.5 & 0.2161 & 0.206--0.226 & 0.1558 & 0.148--0.163 & 0.943 & 0.9--1.2 \\
ECU  & 0.0615 & 0.052--0.072 & 115.5 & 57.8--173.3 & 0.2289 & 0.219--0.239 & 0.0613 & 0.058--0.064 & 0.943 & 0.9--1.2 \\
APL  & 0.0713 & 0.061--0.081 & 56.2 & 28.1--84.3 & 0.1287 & 0.119--0.139 & 0.1270 & 0.121--0.134 & 0.953 & 0.9--1.2 \\
\bottomrule
\end{tabular*}
\end{table*}

Subject-specific \gls{MSK} parameters are optimized during training. 
Table~\ref{tab:optimized_subject4} reports the optimized parameters of a representative subject for all six modeled muscles along with their corresponding physiologically plausible ranges.
All parameters remain within physiologically plausible bounds \cite{Saul2015,zhaoEMGDrivenMusculoskeletalModel2020c}, indicating that \gls{MSK}-NN effectively preserves physiological feasibility.

\subsection{Unmeasured Muscle Ablation Analysis}
\label{subsec:muscle_ablation}
\begin{figure}[t]
    \centering

\begin{subfigure}{\columnwidth}
    \centering
    \includegraphics[width=\columnwidth]{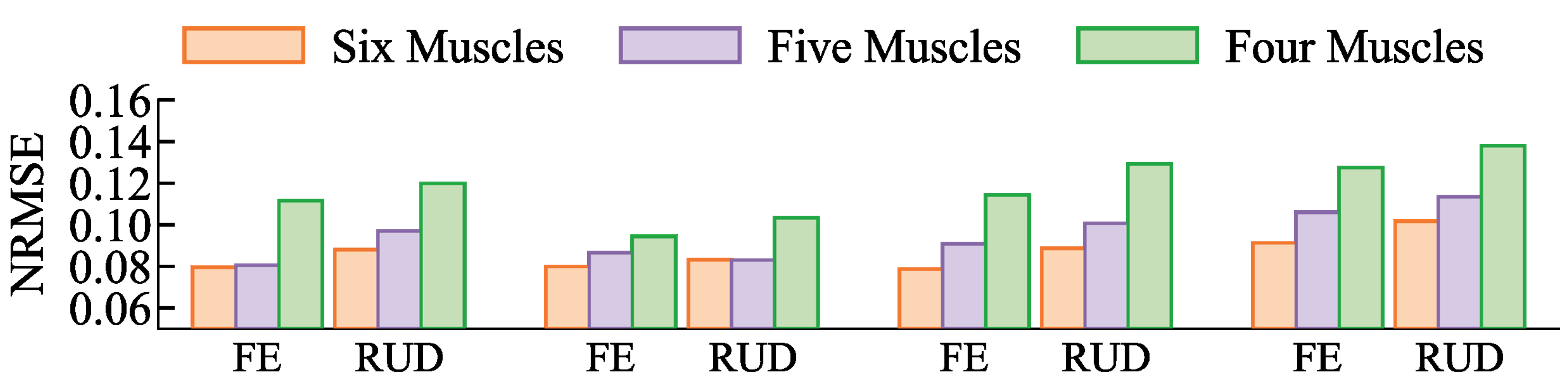}
    
    \vspace{-0.25em}
    \hspace{-0.1\columnwidth}%
    \includegraphics[width=0.98\textwidth]{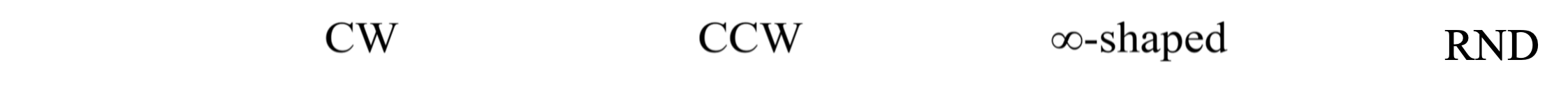}
    \caption{Average \gls{NRMSE} across four wrist motions.}
    \label{fig:reduce_muscle_nrmse}
\end{subfigure}
\vspace{-0.8em}

    \begin{subfigure}{\columnwidth}
        \centering
    \includegraphics[width=\columnwidth]{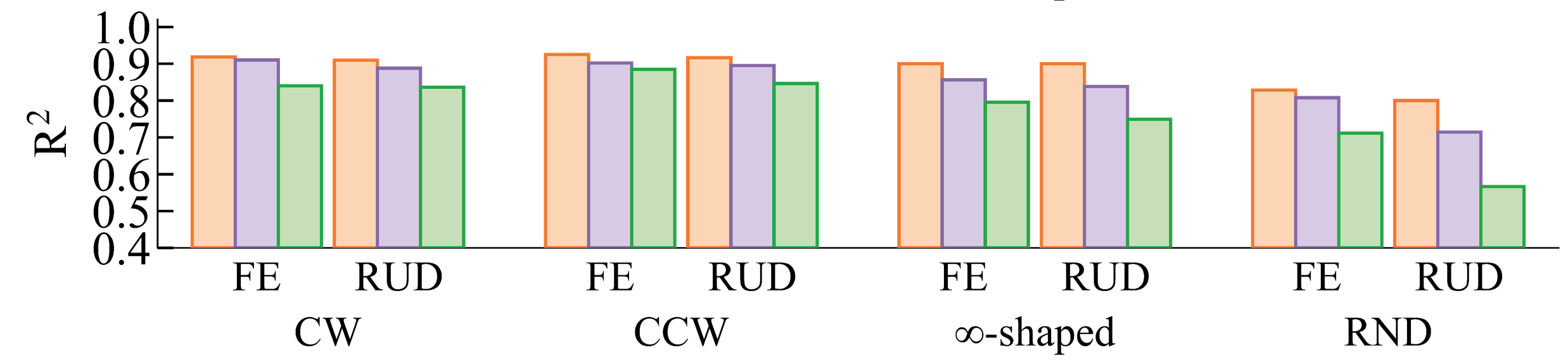}
        \caption{Average \gls{R2} across four wrist motions.}
        \label{fig:reduce_muscle_r2}
    \end{subfigure}

    \caption{
    Muscle ablation analysis across six subjects for three muscle configurations (six, five, and four muscles) across four wrist motions. The five muscle configuration excludes \gls{APL}, and the four muscle configuration additionally excludes \gls{ECRB}.
    }
    \label{fig:reduce_muscle}
\end{figure}

To investigate the contribution of unmeasured muscles within the proposed framework, an ablation analysis is conducted by progressively reducing the muscle set included in the \gls{MSK} forward dynamics module.
Three configurations are compared: the full six-muscle model, a five-muscle model without \gls{APL}, and a four-muscle model without both \gls{APL} and \gls{ECRB}. Each configuration is evaluated across the four motions (\gls{CW}, \gls{CCW}, $\infty$-shaped, and \gls{RND}) for both \gls{FE} and \gls{RUD}.
As shown in Fig.~\ref{fig:reduce_muscle}(a), \gls{NRMSE} increases as muscles are removed, with larger degradation in \gls{RUD} and under the more complex $\infty$-shaped and \glspl{RND}. 
Fig.~\ref{fig:reduce_muscle}(b) shows a corresponding decrease in \gls{R2}, indicating weaker agreement between the estimated and ground-truth joint angles as the muscle set becomes less complete.
Removing \gls{APL} has a limited effect on \gls{FE} estimation but markedly degrades \gls{RUD} performance, consistent with its established role in radial deviation torque generation~\cite{shahImportanceAbductorPollicis2018a}.
Further removing \gls{ECRB} amplifies errors across both \glspl{DoF}, particularly in motions requiring coordinated extensor contributions.

These findings show that including muscles unavailable to the network improves the completeness of the muscle-drive representation and contributes to more accurate multi-\gls{DoF} wrist kinematics estimation under partial \gls{sEMG} observability.

\section{Discussion}
In this section, the proposed \gls{MSK}-NN framework is discussed from three perspectives: modularity and extensibility, its capability to estimate muscle activations under partially observed \gls{sEMG}, and its computational feasibility for real-time execution.

\subsection{Modularity and Extensibility of the Framework}
\gls{MSK}-NN seamlessly integrates two components: a neural activation estimator and an embedded \gls{MSK} forward dynamics module, coupled only through the muscle activation vector that serves as a well-defined physiological interface between the data-driven and physics-based stages.
This compositional structure allows either component to be replaced independently without altering the overall framework.

\gls{MSK}-NN outperforms state-of-the-art baseline methods despite using a lightweight \gls{CNN} architecture for the activation estimator.
For more demanding tasks, this estimator can be replaced with more expressive architectures, such as \gls{TCN} or dual-transformer encoders \cite{Du2024TCNLSTM,wangDualTransformerNetwork2025a}, according to application-specific requirements in data scale, temporal complexity, and latency.
On the biomechanical side, the embedded Hill-type rigid-tendon model can be replaced with higher-fidelity \gls{MSK} formulations.
The physics-physiology loss is architecture-agnostic and can be extended with additional plug-in constraints, such as co-contraction regularization \cite{Wang2025EMGCC}, to satisfy goal-specific objectives.

\subsection{Estimation of Latent Muscle Activations under Partially Observed \gls{sEMG}}
Accurate kinematics estimation under partial \gls{sEMG} observability requires the network to infer activations for muscles unavailable from the input, making the problem underdetermined from kinematic supervision alone.
\gls{MSK}-NN addresses this through two levels of constraints.
At the architecture level, the embedded \gls{MSK} forward dynamics eliminates angle estimations that violate biomechanical causality. 
At the loss level, synergy and trend terms provide soft physiological constraints.
The muscle synergy loss constrains the estimated measured-muscle activations to lie within a \gls{NMF}-derived coordination subspace, thereby encouraging activation patterns that are consistent with low-dimensional muscle coordination structure rather than arbitrary muscle-wise predictions.
The muscle trend loss incorporates anatomical prior knowledge to couple the inference of unmeasured-muscle activations with functionally related measured-muscle activations, allowing information from observed muscles to guide latent activation estimates for muscles unavailable from the \gls{sEMG} input.
The effectiveness of this design is supported by the input-excluded \gls{ECRB} analysis in Section~\ref{subsec:unmeasured} and the muscle ablation results in Section~\ref{subsec:muscle_ablation}.
The agreement between the inferred \gls{ECRB} activation and its recorded \gls{sEMG} envelope suggests physiologically plausible latent activation inference, while the degradation caused by removing \gls{APL} confirms the importance of representing muscles unavailable from the network input.

\subsection{Computational Feasibility for Real-Time Execution}
The proposed \gls{MSK}-NN is evaluated for single-step model latency at 250 Hz on a laptop equipped with an NVIDIA GeForce RTX 4080 Laptop GPU, an Intel Core i9-13980HX processor, and 32 GB RAM.
The CNN-based muscle activation estimator requires 0.55 ms (95th percentile: 0.66 ms), the muscle-tendon dynamics module requires 1.38 ms (95th percentile: 1.60 ms), and the joint dynamics module requires 0.43 ms (95th percentile: 0.51 ms).
The total inference computation time per step is 2.65 ms (95th percentile: 3.17 ms).
These results indicate that the computational core of \gls{MSK}-NN is feasible for real-time execution at 250 Hz.

\section{Conclusion}
In this work, we introduced \gls{MSK}-NN, a physics-embedded neural network for multi-\gls{DoF} joint modeling under partially observed \gls{sEMG}. By integrating a \gls{CNN}-based activation estimator with a differentiable \gls{MSK} forward dynamics module, the framework enables joint kinematics estimation while providing physiologically consistent latent muscle activations.
Experimental results across multiple motions and subjects demonstrate superior kinematic estimation performance compared with baselines and strong temporal agreement between inferred muscle activations and recorded \gls{sEMG} envelopes.
Ablation analysis confirms the importance of both superficial and deep muscles for joint kinematics estimation.
Compared with conventional model-free and hybrid approaches, \gls{MSK}-NN delivers robust performance under sparse sensing while providing interpretable intermediate muscle activations for scientific and engineering applications.
In addition, \gls{MSK}-NN maintains strong performance on random wrist motions, demonstrating robust generalizability.
These findings indicate that the framework can potentially support studies of neuromuscular control under unseen motions, and provide a general template for coupling data-driven estimation with physiologically constrained dynamics.
\bibliographystyle{IEEEtran}
\bibliography{TNRSE}

\end{document}